\pdfoutput=1
\documentclass[a4paper,12pt]{article}
\usepackage{graphicx}
\usepackage{epstopdf, epsfig}
\usepackage{hyperref}
\usepackage{multirow}
\usepackage{booktabs}
\usepackage{setspace}
\usepackage{amsmath}
\usepackage{hyperref}
\usepackage{natbib}
\usepackage{multirow}
\usepackage{epstopdf} 
\usepackage{authblk}
\usepackage{enumerate}
\date{}

\title{\bf{Investigating the correlation between Flat Top Solitary Waves and Offset Bipolar Pulse in auroral plasma}}
\author[1]{S. V. Steffy}
\author[2]{S. S. Ghosh}
\affil[1]{Institute for Integrated Programmes and Research in Basic Sciences, Mahatma Gandhi University,  Kerala, India, 686560.}
\affil[2]{Retired faculty, Indian Institute for Geomagnetism, New Panvel, Navi Mumbai, India, 410218.}

\begin{document}

\maketitle

\begin{abstract}
Non conventional localized pulses, like offset bipolars have been reported in the Earth's magnetosphere. The reports were sporadic and the theoretical interpretations were mainly event based which differ for slow or fast moving pulses. Here it was shown that the offset bipolar pulses are manifestation of an ideal, mathematically defined nonlinear coherent structure called flat top soliton. Adopting Sagdeev pseudopotential technique, and assuming a simple plasma model, it was shown that a slow moving offset bipolar pulse in the auroral region can be interpreted as an ion acoustic flat top solitary wave. The proposed theory is a generic one and can easily be extended to a fast moving one under appropriate plasma model. One important characteristic of such structures is that they generally occur at the boundary of two different phases of nonlinear dynamical processes. 
\end{abstract}

\section{Introduction}
\label{sec:intro}
Electrostatic Solitary Waves (ESWs) are observed ubiquitously in the Earth's magnetosphere. The slow (fast) moving bipolar pulses are often interpreted as ion (electron) mode Solitary Waves (SWs) while the monopolar pulses are usually interpreted as the corresponding Double Layers (DLs). The foundation of such interpretations lie in the theory of coherent nonlinear dynamical structures, like solitons and kink solitons, which are mathematically defined as the solution of a class of integrable Non Linear Partial Differential Equations (NLPDE). High resolution satellite borne instruments, however, identified more kinds of composite and complex organized structures which yet to have a proper theoretical understanding. One such relatively lesser reported structures are the offset bipolar pulses (ofbp) and monopole pairs (mpp) which are the main focus for the present work. The first observational evidence of ESWs was reported in the auroral region by Temerin et al., (\cite{Temerin82}) using S3-3 satellite data. Apart from the conventional bipolar and monopolar structures, they also bore the signatures of ofbp and mpp \citep{Qureshi10}. Witt and Lotko (\cite{Witt83}) theoretically modelled them as paired ion acoustic shocks and showed that they are associated with a square shaped potential well which is different from the conventional bell shaped solitary structures. Their unique theoretical finding, however, didn't seem to have any subsequent follow ups. Later, fast moving offset bipolar pulses were observed in the day side Polar Cap Boundary Layer (PCBL) (\citep{Tsurutani98}), the downward current regions of the auroral zone (\citep{Muschietti01}, \citep{Roth02}), the diffusion region of reconnection (\citep{Deng05}), and in the magnetosheath (\citep{Pickett08}) by POLAR, FAST, GEOTAIL, and CLUSTER satellites. It was  Tsurutani et al.,\citep{Tsurutani98} who identified them as unique kinds of coherent localized structure. Following their idea, and incorporating a Bernstein, Greene and Kruskal (BGK) model \citep{Muschietti01} and Particle In Cell (PIC) simulation \citep{Roth02}, ofbps were interpreted as flat shaped electron phase space holes, emerging due to the trapping of electrons. Such a theory, however, remain inadequate to explain slow moving ion mode ofbps. 

In spite of these sporadic reports, and a strong indication that they may appear consistently and regularly at different parts of the Earth's magnetosphere, the theory of the offset bipolar structures received a comparatively scant attention. Even those limited attempts mentioned so far were more focused to interpret a specific set of observations rather than developing a generic theory for ofbp. This lacuna motivated us to explore a more generic interpretation for ofbp which would be consistent with the observed "stretched bipolar structures", alternately known as offset bipolar pulses (ofbp), during satellite expeditions. 

Both the electron mode BGK models, and the ion acoustic ofbp based on ion fluids, show that they are associated with a square shaped, or "flat top" solitary profile. During our theoretical analysis, we have indeed obtained such Flat Top Solitary Wave (FTSW) solutions with square shaped potential profiles. The nomenclature not only describes the unique morphology of the potential profile but also rightly connects it to a more analytical and ideal set of solutions called Flat Top Solitons (FTS). Analogous to a solitary wave vis \'a vis a soliton solution, an FTSW is the realistic counter part of the FTS which is obtained by using Sagdeev pseudopotential technique \citep{Steffy18b},\citep{Verheest20}. The technique is widely used to predict the existence and characteristics of any nonlinear coherent structure by studying the trajectory of the pseudoparticle in the pseudopotential well without the rigorous solution of any particular NLPDE, as is essential for a soliton, or FTS solution.

Previously, the theory of FTS solutions has been successfully applied in photonics \citep{Mayteevarunyoo08} and internal ocean waves \citep{Jeans01,Stanton98}. Here, for the first time, we have extended the same idea to space plasma for interpreting the ofbp. Rather than being event based, the theory proposes a more generic approach which correlates the observed ofbp with a more fundamental nonlinear dynamical structure called FTS through FTSW where the latter retains all the physical characteristics and the boundary conditions of FTS. We further conjecture that the structure is an amalgamation of a SW and a DL and may appear at the boundary of two different phases of nonlinear dynamical processes.

The paper is organized as follows. Assuming a simple plasma model, in Sec 2 we have found an ion acoustic ofbp which we have further validated with the corresponding satellite observations. In Sec. 3 we have explored the generic characteristics of the associated FTSW which is the steady state generalization of an FTS. The concluding remarks are given in Sec. 4

\section{Validation of FTSW with space plasma observations}
Slow moving ESWs, moving with ion acoustic speed, have been observed in the low altitude auroral region by several spacecraft expeditions, like S3-3 \citep{Temerin82}, VIKING\citep{Bostrom89}, POLAR \citep{Bounds99}, and FAST\citep{McFadden03} satellites. S3-3 recorded slow moving ofbp at an altitude between $6000\,km$ and $8000\,km$ \citep{Qureshi10}.  At this altitude, the plasma has been found to have a significant contribution of $O^{+}$ ions along with its usual proton ($H^{+}$) population. Moreover, there is an admixing of hot magnetospheric electrons with a cooler component originated from the ionosphere.  The tenuous plasma condition and the absence of any physical boundary allow us to assume the plasma to be collisionless, homogeneous, and infinite. Following Temerin et al., \citep{Temerin82} we have further assumed that the wave is moving along the ambient magnetic field, making the plasma isotropic and unmagnetized. It is observed that the speed of the ESW in this region is of the order of the ion acoustic speed of the medium, with the wave speed $V\approx50~km/s$, which indicates that they are governed by the ion dynamics. To sustain any such wave, the electron temperature should be higher than the corresponding ion temperatures. Besides, they have negligible inertia compared to ions. Theoretically it is well known that a secondary component of electrons is necessary to sustain an ion acoustic DL. The same condition was found to hold true for a Super Solitary Wave (SSW)\citep{Ghosh14, Steffy16}, or an FTSW\citep{Steffy18b} as well.  Hence we have assumed that the plasma has two electron temperatures, both obeying Boltzmann distributions and are separately in thermal equilibrium. The overall plasma is a four component one with warm multi-ion fluids comprising $H^{+}$ and $O^{+}$ ions so that the corresponding mass ratio $Q=1/16$.

The recorded ambient plasma density of this region is of the order of $n_0\approx5-10\, cm^3$. Following Temerin et al., \citep{Temerin82}, we have chosen an ambient plasma density $n_0=10\,cm^3$, giving rise to an overall proton plasma frequency for $n_0$ as $\omega_{pi}=4.163\,kHz$. We have assumed a very low concentration of cooler electrons (viz., $0.12\%$ of $n_0$) and sufficiently small presence of $O^{+}$ ions ($10\%$ of $n_0$), leading to the normalized ambient densities  $\mu = 0.0012$ for cooler electrons and $\alpha_h = 0.1$ for $O^{+}$ ions, respectively. All the number densities were normalized by the ambient plasma density $n_0$. Satellite observations have recorded cooler electron temperature $T_{ec}\approx 0.5 - 5\, eV$ \citep{Temerin82},\citep{kosk}. Hence, in this work for our convenience, we have chosen $T_{ec}=0.5\, eV$, and the electron temperature ratio $\tau=\frac{T_{ec}}{T_{ew}}=0.0485$ so that $T_{ew}=10.3\,eV$. Since both the electrons are taking part in the Debye shielding, we have estimated the effective temperature  $T_{eff}=\frac{T_{ec}T_{ew}}{\mu T_{ew}+(1-\mu) T_{ec}}= 10.0722\,eV$ which further gives us the estimated effective Debye length $\lambda_{d_{eff}}=7.4606\,m$  and the effective ion acoustic speed for protons $c_{isl}=31.06\,km/s$. These three parameters, together with $\omega_{pi}$, determine the overall scale of our plasma system. This is consistent with the observations of Temerin et al. \citep{Temerin82} who have reported a Debye length of $\lambda_d \approx 5\,m$ for the said region. For our theoretical analysis, we have normalized all the space variables by $\lambda_{d_{eff}}$, time by  $\omega_{pi}^{-1}$, temperatures by $T_{eff}$, and the electrostatic potential $\phi$ by $T_{eff}/e$. All the speeds, along with the wave Mach number $M$, are normalized by the $c_{isl}$.

In the literature, there was a mention of the cooler ions with the ion temperature of the order of a few $eV$ \citep{kosk}. Hence in order to sustain the ion acoustic wave the ions are necessarily cooler than electrons, we have chosen the lighter ion temperature $T_{il}=0.33\, eV$ which is cooler than the cooler electron temperature. Theoretically we have found that the ion temperature plays a marginal role in determining an FTSW solution compared to the corresponding electronic parameters, such as $\mu$ and $\tau$. Besides, the effect of $O^+$ ions is expected to be even smaller because of its apparently low concentration. In the absence of any clear mention of the type the ion species for the hotter one, and for our analytical convenience, we have assumed that both the $H^+$ and $O^+$ ions have equal temperatures, giving rise to an overall normalized ion temperature $\sigma=0.033$, normalized by $T_{eff}$.

 To ensure a steady state condition, or a wave frame, we have further assumed $\eta$ to be the generalized coordinate where
\begin{equation}
\eta=x-Mt
\end{equation}
The corresponding Sagdeev pseudopotential $\Psi(\Phi)$ for the chosen plasma model is \citep{Steffy16}
\begin{equation}
\begin{split}
\psi(\Phi)= -\Bigg[\bigg\{\mu+(1-\mu)\tau\bigg\}\bigg\{\mu\bigg(exp \frac{\Phi}{\mu+(1-\mu)\tau}-1\bigg)+\frac{1-\mu}{\tau}\bigg(exp \frac{\tau\Phi}{\mu+(1-\mu)\tau}-1\bigg)\bigg\}+\\
\frac{\alpha_l}{6\sqrt{3\sigma_l}}\bigg\{[(M+\sqrt{3\sigma_l})^2-2\Phi]^{\frac{3}{2}}-(M+\sqrt{3\sigma_l})^3-\\ [(M-\sqrt{3\sigma_l})^2-2\Phi]^{\frac{3}{2}}+(M-\sqrt{3\sigma_l})^3\bigg\}+\\
\frac{\alpha_h}{6\sqrt{3\sigma_h}}\bigg\{[(\frac{M}{\sqrt{Q}}+\sqrt{3\sigma_h})^2-2\Phi]^{\frac{3}{2}}-(\frac{M}{\sqrt{Q}}+\sqrt{3\sigma_h})^3-\\
[(\frac{M}{\sqrt{Q}}-\sqrt{3\sigma_h})^2-2\Phi]^{\frac{3}{2}}+(\frac{M}{\sqrt{Q}}-\sqrt{3\sigma_h})^3\bigg\}\Bigg]
\end{split}
\end{equation}
which satisfies the following 'energy equation' \citep{Steffy16}

\begin{equation}
\frac{1}{2}\Big(\frac{d\Phi}{d\eta}\Big)^2+\Psi\left(\Phi\right)=0
\end{equation}

\noindent Eq. (2.3) is a modified form of the Poisson's equation where the slope of the pseudopotential is defined as the associated charge separation, i.e., $\frac{\partial\Psi(\phi)}{\partial\Phi}=\Delta n(\phi)$ for any $\phi$, $\Delta n=n_i-n_e$ being the charge separation between overall ion $(n_i)$ and electron $(n_e)$ charge densities . A $\Psi$ Vs $\Phi$ curve determines the existence of the localized nonlinear coherent structures, like SWs, SSW, or FTSW, provided following boundary conditions are satisfied
%\begin{linenomath*}
\begin{subequations}
\begin{align}
\Psi(\Phi=0)&=\frac{\partial\Psi}{\partial\Phi}\Big|_0=0;\;\;\;\frac{\partial^2\Psi(0)}{\partial\Phi^2}<0\\
\Psi(\Phi_0)&=0;\;\;\;\frac{\partial\Psi(\Phi_0)}{\partial\Phi}\neq0;\;\;\;\Psi(\Phi)<0 \;\, {\rm for}\;\, 0\geq\Phi\geq\Phi_0.
\end{align}
\end{subequations}
%\end{*}

\noindent where $\Phi_0$ is the amplitude of the wave structure.

The last condition in Eq. (2.4) ensures the recurrence of the initial state for a soliton, or SW. For a DL, this condition modifies as
%\begin{linenomath*}
\begin{equation}
\frac{\partial\Psi(\Phi_0)}{\partial\Phi}=\Delta n_d=0
\end{equation}
%\end{linenomath*}
where $\Delta n_d$ is the charge separation at the maximum amplitude for the DL. Studying the trajectory of the "pseudoparticle" in the said pseudopotential well, and implementing the above mentioned 'boundary conditions' for the pseudoparticle, viz., Eqs. 2.4 and 2.5, the Sagdeev pseudopotential technique enables one to predict the presence of the corresponding steady state nonlinear structure.

\begin{figure}
\centering
\includegraphics[scale=0.8 ]{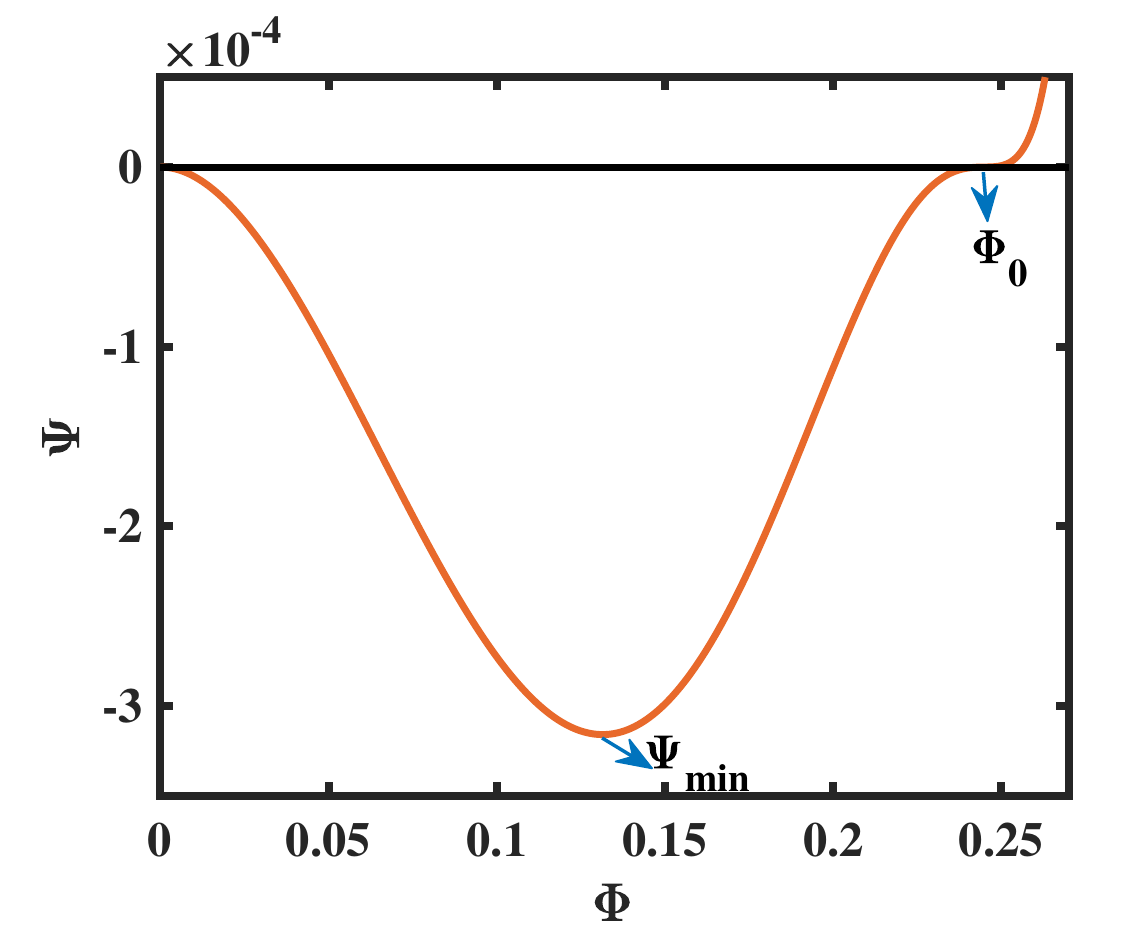}
\caption{Sagdeev pseudopotential profile corresponding to an FTSW.}
\label{figure1}
\end{figure}

For $M=1.05663685$, and for the chosen set of parameters as mentioned above, we have obtained the Sagdeev pseudopotential curve as shown in Fig. \ref{figure1}. Since it satisfies Eq. 2.4, it represents a SW which is a steady state analog of a soliton. 

To find the corresponding electric field profile, we have deduced the generalized electric field $E = -\frac{d\Phi}{d\eta}$ from Eq. (2.3). The time duration $\Delta t$ (non-normalized) is estimated from from the half-width $W$ assuming $x=0$ in Eq. (2.1). The half-width $W$ is defined from Eq. (2.3) as

\begin{equation}
W=2\eta_{1/2}=2\int_{\Phi_0}^{\frac{\Phi_0}{2}}\Bigg(\frac{1}{\sqrt{(-2\Psi(\Phi))}}\Bigg)d\Phi\,;\;\; \Delta t=\Bigg(\frac{W}{M}\Bigg)^{\prime}\;\, {\rm for}\;\, x=0
\end{equation}

\begin{figure}
\centering
\includegraphics[scale=0.8 ]{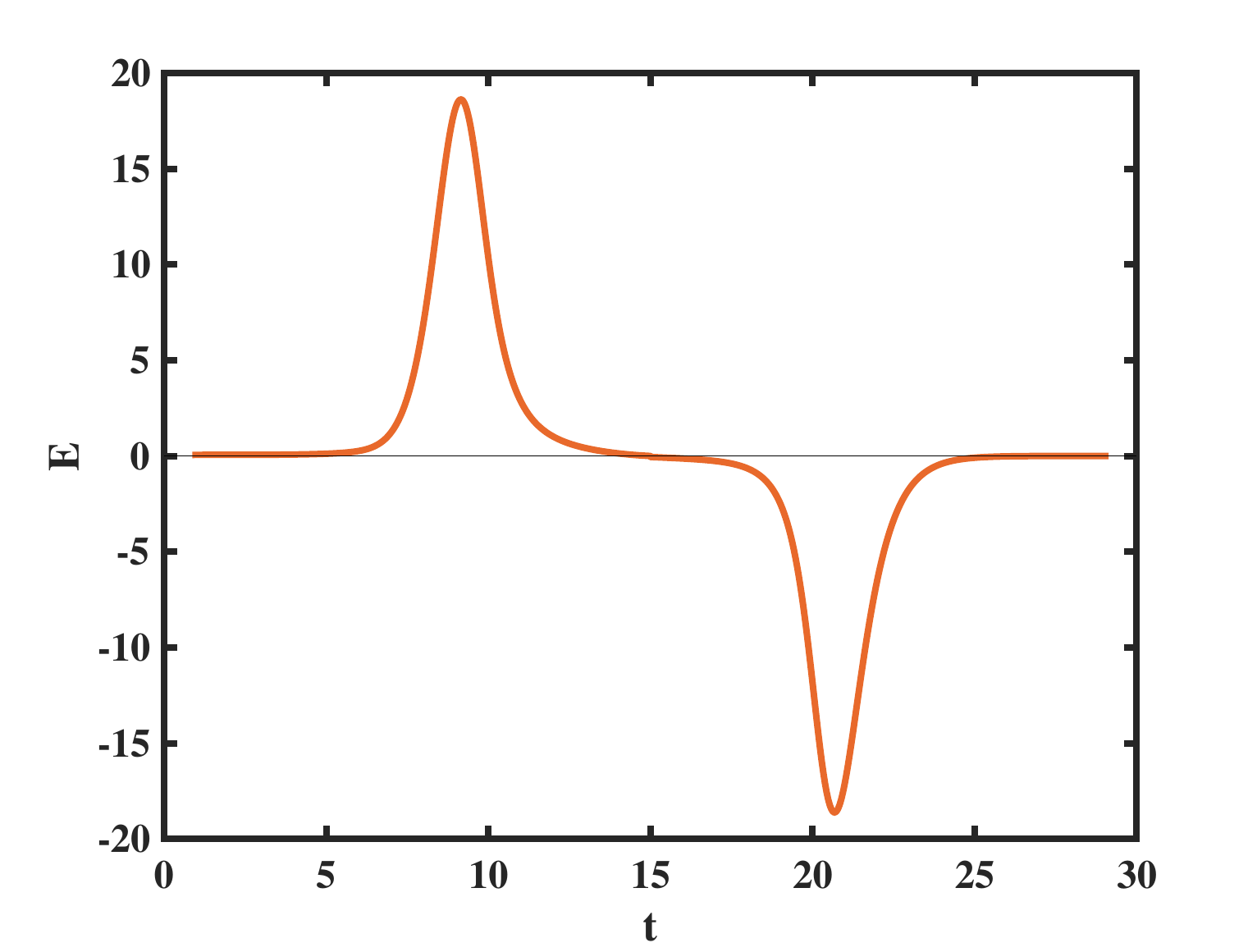}
\caption{Non-normalized electric field (E) profile corresponding to an FTSW.}
\label{figure2}
\end{figure}

\noindent The normalized half-width $W=422.55\,m$ and the prime ($\prime$) in Eq. 2.6 denotes the corresponding non-normalized parameters. Figure \ref{figure2} shows the associated electric field profile in non-normalized parameters which clearly reveals an ofbp. We have further estimated the average ($E_{avg}=\big(\frac{\Phi_0}{W}\big)^{\prime}$) and the maximum ($E_{max}= -\big(\frac{d\Phi}{d\eta}\big)^{\prime}\big|_{\Psi=\Psi_{min}}$) electric fields analytically, where $\Phi_0=0.2448$ is the normalized potential amplitude and $\Psi_{min}$ is the minimum value of $\psi$ for $0\leq\phi\leq\phi_0$, both of them are marked by the respective arrows in Fig. \ref{figure1}. The estimated average electric field across the structure $E_{avg}=5.8387\,mV/m$, and the estimated time duration $\Delta t=12.8746\,ms$ (Fig.\ref{figure2} ). The S3-3 satellite observations have revealed an $E \leq 15\, mV/m$ and $\Delta t=2-20\, ms$ which are a close match to our analytical estimations. The analytically estimated peak to peak E-field amplitude $E_{p-p}=2\left|E_{max}\right|\approx 37.02\, mV/m$, have higher amplitude than the observed E-field. This still remain consistent with the qualitative agreement since a fluid approximation is known to overestimate the amplitide \citep{Ghosh98}. We have further estimated the speed of the wave structure $V=32.83~km/s$ from our chosen Mach number $M(=1.05663685)$ which is in accordance with the satellite observations, i.e.,$V\approx50~km/s$, as mentioned earlier. Table 1 compares shape, size, and speed of an ESW in the auroral region with those estimated analytically for a possible ofbp. The latter shows a comparatively wider $\Delta t$, as expected from its stretched structure. Our results indicates that the ofbp obtained analytically here using Sagdeev pseudopotential is a feasible candidate to interpret the corresponding satellite observations.  A more rigorous modelling of ofbp with a complete and contemporary data is beyond the scope of the present work and may be presented elsewhere.

\begin{table}
\centering
\caption{Wave parameters}
\label{two}
\begin{tabular}{|c|c|}
\hline
Theoretical    &Observational \\ \hline%\cline{1-2}
%Normalized  & Non-normalized          &  \\ \hline
%$\Phi_0=0.2248$ &  
$E_{avg}=5.8387~mV/m$   & $E\leq15~mV/m$                                   \\ \hline
%$W=603.2$     &  
$\Delta t=12.8746~ms$     &    $\Delta t=2-20~ms$                                     \\ \hline
 %               &
 $V=32.82~km/s$  &  $V\approx50~km/s$                                 \\ \hline
\end{tabular}
\end{table}

% susie put cite commands here, don't bother with citet etc just yet.

\section{Physical properties of FTSW and ofbp}
\begin{figure}
\centering
\includegraphics[scale=0.8 ]{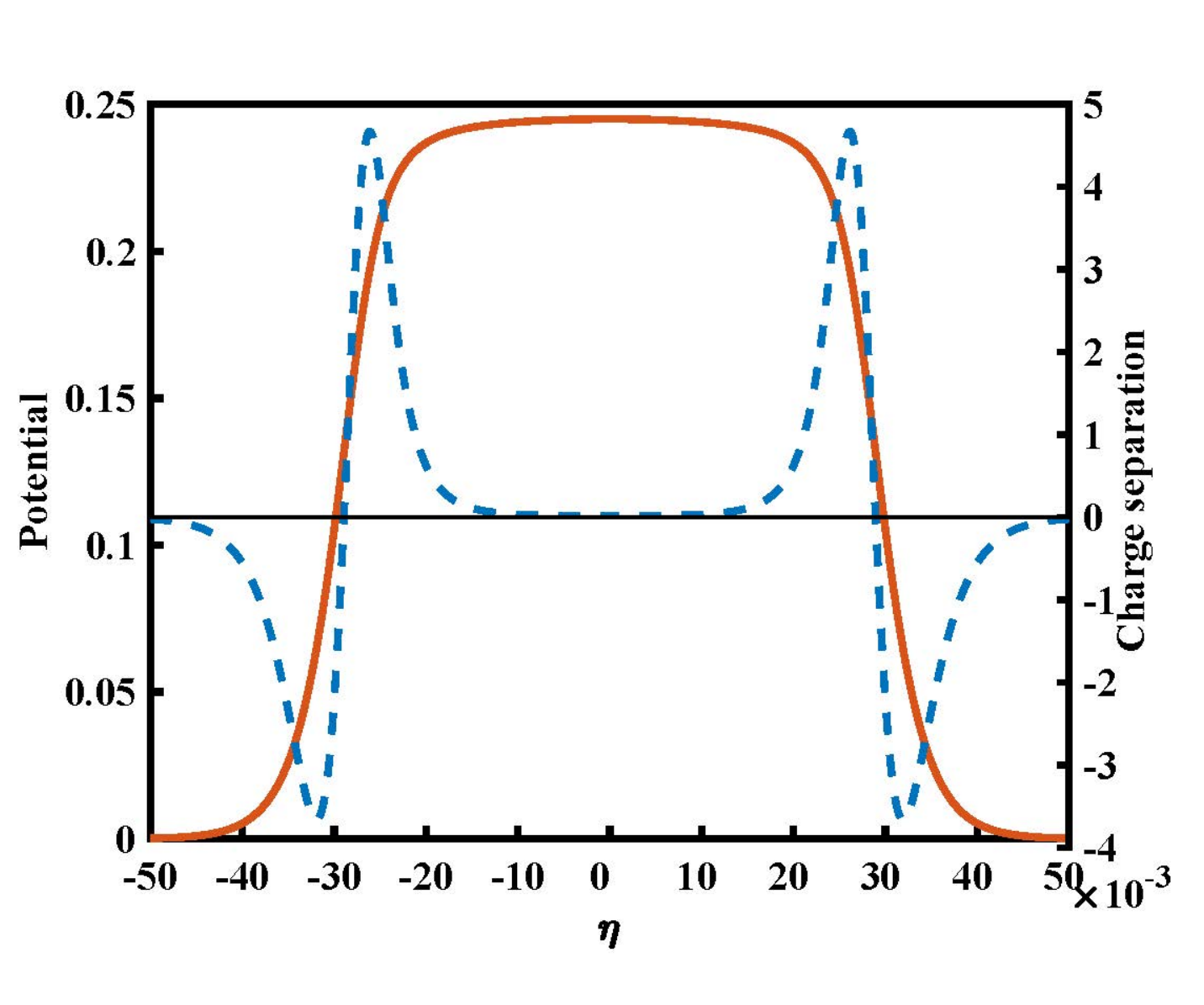}
\caption{Potential profile (solid line) and charge separation profile (dashed line) corresponding to an FTSW.}
\label{figure3}
\end{figure}

To explore the physical characteristics of an ofbp, we have plotted the corresponding potential (solid line) and charge separation (dotted line) profiles in in Fig.\ref{figure3}. It confirms a flat top profile for an FTSW. Conceptually, an FTSW can be visualized as an amalgamation of a SW and a DL. Its charge separation near the maximum amplitude turns vanishingly small (dotted line), approaching the condition of a DL (Eq. (2.5)), although it always remain finite and non-zero, eventually satisfying Eq. (2.4) and the solution bounces back to its initial state like any conventional soliton. The modified condition for an FTSW can thus be written as

\begin{equation}
\Psi(\Phi_0)=0\,;\;\;\;\frac{\partial\Psi}{\partial\Phi}\Big|_{\Phi_0}=\epsilon,\;\frac{\partial^2\Psi}{\partial\Phi^2}\Big|_{\Phi_0}=\delta\,;\;\;\;\epsilon,\delta\neq0\,; %\left|\epsilon\right|,\,\left|\delta\right|\ll 1
\end{equation}

\noindent where $\epsilon$, $\delta$ are two arbitrarily small, but finite numbers. The grazing incidence (i.e., low slope) of the curve to the $\Phi$ axis at $\Phi_0$ satisfies Eq. (3.1) for an FTSW while its finite slope at $\Phi_0$ satisfies Eq. (2.4) as well. This implies that the 'pseudoparticle', associated with the solution, leaves $\Phi=0$ at rest and reaches its reflection point at $\Phi_0$ after a prolonged time, as ascertained by the grazing incidence of $\Psi$, and  then it oscillates back to $\Phi=0$ giving rise to an wider, but well localized, coherent structure similar to a conventional SW. The morphology of the structure, however, is different from that of a conventional one as is evident from the associated electric field and potential profiles.

Previously Roth et al. (\citep{Roth02}) achieved an ofbp by assuming flat top potential profile for their PIC simulation. In the present case, the FTSW and the associated ofbp have been obtained analytically from a simple plasma model  without any prior assumption of the potential profile. It was previously indicated that the trapping of electrons in BGK phase space hole may cause an ofbp. Such an assumption will not be valid for an ion mode, positive amplitude ofbp. However, in spite of the differences in the respective plasma models and associated techniques, the charge separation profiles remain the same for both the cases. We have previously observed that, when a Regular Solitary Wave (RSW) transits to a DL in the parameter space, its charge separation at its maximum amplitude drops, becoming ideally zero for a DL. We can here visualize an FTSW as an 'incomplete DL' where the solution goes very close to a DL solution but stops short of it and bounces back retaining the characteristics of a SW along with a strong imprint of a DL-like solution as well. This is manifested in the long sustentation of the vanishingly small charge separation which eventually causes the flat top profile. It is the same characteristic which causes the stretching in the localized E-field making it offset bipolar.

It is now evident that the ofbp, and the FTSW, both indicate the same localized coherent structure. The concept now can be extended further to the more ideal solutions called FTS which, as we have discussed earlier, is the mathematical counterpart of an FTSW, in the same way as a soliton is correlated with a more general class of solutions we call as solitary waves. It is well known that the integrable NLPDEs, like Korteweg-de Vries (KdV), govern soliton solutions. Analogously, a modified, or extended form of KdV (eKdV), popularly often known as the Gardner equation, governs FTS solutions depending on its specific boundary conditions \citep{Grimshaw02}. Like eKdV, there are other such modified NLPDEs which also have FTS as one of their possible solutions. As per previous literature, an ofbp has often been described as a 'stretched' or 'dispersed' bipolar pulse where it was conjectured that the stretching has been happened due to an extra dispersion in the medium. The solution obtained from an e-KdV or Gardner equation describes the significance of an extra cubic nonlinearity term which results in a secondary insurgence in the nonlinearity” that balances the excessive dispersion maintaining the solitary structure.

\begin{figure}
\centering
\includegraphics[scale=0.8 ]{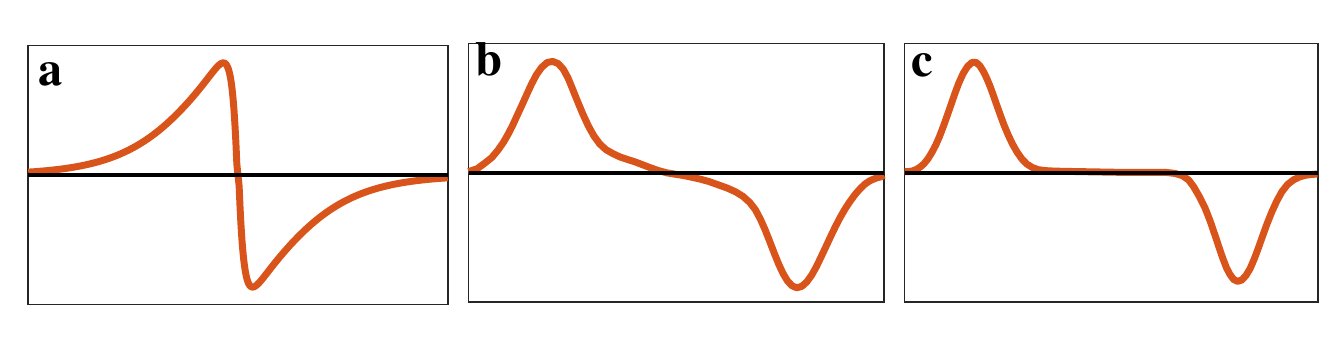}
\caption{Schematic diagram of (a) bipolar E-field pulse (b) offset bipolar E-field pulse and (c) paired monopolar E-field pulse}
\label{figure4}
\end{figure}

To complement our understanding on ofbp, here we recall a schematic diagram in Fig.\ref{figure4} which explains the characteristics of an ofbp (Fig. \ref{figure4} b) vis \'a vis a conventional bipolar and and a monopole pair (mpp) in Figs. \ref{figure4}a  and \ref{figure4}c respectively. For both an ofbp and an mpp, the distances between the two peaks are relatively large compared to the characteristic width of the each peak. The fine difference between an ofbp and mpp lies in the finite slope of the E-field connecting the two lobes for the former which categorizes it as a solitary wave in general. An mpp, on the other hand, is a pair of two simple monopoles with opposite polarities (Fig. \ref{figure4}c) while the slope of the electric field connecting the two poles goes ideally to zero. The significance of the finite slope in E reflects the low, but non zero slope of the Sagdeev pseudopotential at its maximum amplitude which in turn defines the respective boundary conditions of different nonlinear structures, viz., SW (Eq. 2.4), DL (Eq. 2.5), and FTSW (Eq. 3.1). This further ensures an unique solution for $\Phi_0$, the maximum amplitude, while an ideally 'flat' solution would be analytically ambiguous due to its lack of uniqueness. Besides providing a physical explanation for the morphology of an ofbp, it further enhances the candidature of the FTSW to interpret the observed structures.

Recently Qureshi et al. \citep{Qureshi10} generalized the concept of ofbp beyond the acoustic mode as they predicted both ion acoustic and ion cyclotron ofbp for their theoretical model. Though their results are yet to be validated by the observational data, it indicates that the structure is more generic than it was thought so far. According to the present understanding, the offset bipolars are appearing like a sporadic "deformation" of the conventional bipolar electric field, resulting due to certain arbitrary local conditions at the spot. The proposed theory of FTSW, on the other hand, generalizes an ofbp beyond its local conditions correlating it to an FTS which is, like soliton, is known to exist across different physical situations, even beyond the realm of the plasma physics. Apart from its unique morphology, one common characteristic of an FTS is that they often define a boundary between two phases or nonlinear dynamical processes. Similar characteristic has also been reported for an FTSW as well. During our theoretical analysis, we have found that the FTSW is occurring at the boundary between two types of SSWs, viz. Type I and Type II, where an SSW is characterized by the extra wiggles in their otherwise bipolar electric field \citep{Steffy18b}. A Type I SSW associates it with a preceding DL / monopole while a Type II emerges due to a continuous deformation of the bipolar electric field. We may here conjecture that, analogous to an FTSW or FTS, an ofbp, too, may define a boundary between two phases, or nonlinear dynamical processes in the space. A more rigorous mathematical derivation of an ofbp from the preliminary FTS solution is beyond the scope of the current paper and may be presented elsewhere.

\section{Conclusion}

Using a simple plasma model, we have analytically estimated the shape, size, and speed of an FTSW which was found to be consistent with the slow moving ofbp observed in the Earth's auroral region.  It manifests that ofbps are eventually FTSWs, or even may be FTS, where the latter is the more mathematical and ideal counterpart of FTSW. Following the theory of the FTSW and FTS, we have interpreted the ofbp as an amalgamation of SW and DL which determines the boundary of two distinct phases of nonlinear dynamical processes. The proposed theory not only explains the unique morphology of the E-field data but also provide a more generic interpretation for the ofbp which is eventually correlating the mathematical description of a coherent nonlinear dynamical structure with the satellite observations. It is expected to provide a new way of understanding the non conventional localized pulses in the E-field data recorded during satellite expeditions which are known to be important in determining the microphysics of the Earth's magnetospheric boundary layers.

\section{Acknowledgement}

The authors would like to thank the Director, Indian Institute of Geomagnetism, Mumbai for providing the necessary facilities for carrying out the research work.

\bibliographystyle{jpp}

\bibliography{steffy_draft}

\end{document}